\documentclass[pre,amssymb,amsmath,twocolumn,aps,showpacs]{revtex4}
\usepackage{graphicx}
\usepackage{dsfont}

\usepackage[dvips]{epsfig}
\usepackage[usenames]{color}

\newcommand{\rev}{\bar}
\begin{document}
\title{Echo states for detailed fluctuation theorems}
\author{T. Becker}
\email{thijsbecker@gmail.com}
\affiliation{Hasselt University, B-3590 Diepenbeek, Belgium}
\author{T. Willaert}
\affiliation{Hasselt University, B-3590 Diepenbeek, Belgium}
\author{B. Cleuren}
\affiliation{Hasselt University, B-3590 Diepenbeek, Belgium}
\author{C. Van den Broeck}
\affiliation{Hasselt University, B-3590 Diepenbeek, Belgium}

\date{\today}

\begin{abstract}
Detailed fluctuation theorems are statements about the probability distribution for the stochastic entropy production along a trajectory. It involves the consideration of a suitably transformed dynamics, such as the time reversed, the adjoint, or a combination of these. We identify specific, typically unique, initial conditions, called echo states, for  which the final probability distribution of the transformed dynamics reproduces the initial distribution. In this case the detailed fluctuation theorems relate the stochastic entropy production of the direct process to that of the transformed one. We illustrate our results by an explicit analytical calculation and numerical simulations for a modulated two-state quantum dot.
\end{abstract}
\maketitle
\section{Introduction}
The discovery of detailed and integral fluctuation theorems is arguably one of the most significant recent advances in nonequilibrium statistical mechanics \cite{AIPevans,RMPesposito,RPPseifert}. The best known examples are the Jarzynski equality \cite{PRLJarzynski} (integral fluctuation theorem) and the Crooks relation~\cite{PRECrooks2008} (detailed fluctuation theorem).
The Crooks relation implies the consideration of the time-reversed dynamics. As was already pointed out by Crooks himself~\cite{PRECrooks2008,noteSeifert}, the application of the theorem involves a stringent condition on the initial condition of the reverse process, namely, that it be such that the final distribution of this reverse process is the initial distribution of the forward process; see also \cite{PREverley,JSMharris2007}. In that sense, the integral fluctuation theorems appear to have a broader range of validity, a point made particularly clear in the work by Speck and Seifert \cite{JPAspeck2005,JSMspeck2007}. 

The  fluctuation theorems derive from time-symmetry properties of the underlying microscopic dynamics. It was however realized that one can, at least in the context of Markovian processes, consider two types of  symmetry operations related to time-irreversible behavior. Besides the time inversion of the driving, one can perform the time inversion of the dynamics associated to nonequilibrium boundary conditions. This is technically done by considering the  adjoint of the Markov operator. It was thus found by Esposito and Van den Broeck \cite{PRLesposito2010} that there are three different types of integral and detailed fluctuation theorems. Each of these fluctuation theorems is associated to one of the three combinations of the two symmetry operations, corresponding to the total entropy production, the non-adiabatic entropy production (loosely speaking associated to relaxation processes), or the adiabatic entropy production (loosely speaking related to the dissipation of  the nonequilibrium steady states).  This discovery clarified the status of  the Speck-Seifert and Hatano-Sasa \cite{PRLhatano2001,JSMchernyak2006} fluctuation theorems and of the $H$ theorem familiar from the theory of stochastic processes \cite{PREesposito2010}. In the adiabatic fluctuation theorem, the same initial condition is considered for both the original and transformed processes. For the total and non-adiabatic fluctuation theorem the transformed process starts with the final distribution reached in  the forward process. The question can be raised whether the final distribution of this transformed process reproduces the initial distribution of the forward process, as this has a direct consequence on the interpretation of the fluctuation theorems. The main purpose of this paper is to show that there is indeed a, generically unique, initial distribution for which this property holds.  We call such an initial distribution an ``echo state." As we will see, the echo state depends in an intricate manner on the details of the dynamics and is, in general, different for the total and non-adiabatic entropy productions.

The identification of the echo states is of particular interest in the case of time-periodic driving, which we discuss in more detail. In the experiments considered so far, the measurements were restricted to the case in which the steady state was, for symmetry reasons, trivially identical to the echo state \cite{PRLSeifert,PRLBlickle2006,EPLjoubaud2008}. This need not be the case. We will show how to identify the echo states in a general setting, and how to extract the proper statistics even when not operating with an echo state as the initial condition, by an appropriate ``shadowing operation." Finally, we illustrate how our prescriptions can be implemented with the analytic and numerical discussion of a modulated two-state quantum dot. 

\section{Notation and definitions}\label{1}
We consider systems with Markovian dynamics and a discrete space of states. 
The transition rates from a state $m'$ to a state $m$ are denoted by $W^{(\nu)}_{m,m'}(\lambda_t)$, where $\lambda_t$ is a time-dependent control variable that describes the external driving and $\nu$ specifies the mechanism that causes the transition. Transitions are caused by contact with equilibrium reservoirs. The total transition rate from $m'$ to $m$ is denoted by $W_{m,m'}(\lambda_t) = \sum_{\nu} W^{(\nu)}_{m,m'}(\lambda_t)$. 
The probability to be in state $m$ at time $t$ obeys the following master equation:
\begin{equation}
\dot{p}_m(t) = \sum_{m'} W_{m,m'} (\lambda_t) p_{m'}(t),
\end{equation}
or $\dot{\mathbf{p}}_t = \mathbf{W}(\lambda_t) \mathbf{p}_t$ in vector notation. The initial condition is given by the probability distribution $\mathbf{p}_0$. The diagonal elements of the rate matrix satisfy $W_{m,m}(\lambda_t) = -\sum_{m' \neq m} W_{m',m}(\lambda_t)$.
 In the absence of driving ($\lambda_t=\lambda$), a system in contact with a single reservoir $\nu$ will relax to the equilibrium distribution $\mathbf{p}^{\mathrm{eq},(\nu)}(\lambda)$. When the system is in contact with multiple reservoirs, it will relax to a nonequilibrium steady state (NESS) $\mathbf{p}^{\mathrm{s}}(\lambda)$.
The time evolution of the system is described by a trajectory $\Pi = \{m(t), t \in [0, T]\}$.
The time of the $i$th jump is denoted by $t_i$ ($1 \leq i \leq N$), with $N$ the number of jumps. The trajectory starts at $t_0 = 0$ and ends at $t_{N+1} = T$. A trajectory is completely specified by its jump times $t_i$, state prior to the jump $m_{i-1}$, state after the jump $m_i$, and reservoir that causes the jump $\nu_i$.
The probability to observe a trajectory $\Pi$, given $\mathbf{p}_0$, is equal to
\begin{multline}
\mathcal{P}(\Pi  \vert \mathbf{p}_0)= p_{m_0}(0) e^{\int_{t_N}^{T} d \tau W_{m_{N},m_{N}}(\lambda_{\tau})}\\
\times \left[ \prod_{j=1}^N e^{\int_{t_{j-1}}^{t_j} d \tau W_{m_{j-1},m_{j-1}}(\lambda_{\tau})} W^{(\nu_j)}_{m_j,m_{j-1}}(\lambda_{t_j}) \right].\label{eq::probpathfwd}
\end{multline}
In order to define the total trajectory entropy production (EP) we need to introduce the time-reversed trajectory $\rev{\Pi} \equiv \{ \rev{m}(t) =  m(T-t),  t \in [ 0, T] \}$. Also, with the (forward) driving $\lambda_t$ we can associate the time-reversed driving $\rev{\lambda}_t \equiv \lambda_{T-t}$. 
For later reference, we introduce the following (stochastic) matrices:
\begin{equation}\label{eq::WFWR}
\mathbf{W}_{\mathrm{F}} =  \overrightarrow{\exp} \int_0^T\!\!\!\!\mathbf{W}(\lambda_t) dt\;\;\;;\;\;\;
\mathbf{W}_{\mathrm{R}} = \overrightarrow{\exp} \int_0^T \!\!\!\!\mathbf{W}(\rev{\lambda}_t) dt,
\end{equation}
where $\overrightarrow{\exp}$ stands for the time-ordered exponential. These matrices describe the time evolution from $t=0$ to $t=T$ of, respectively, the forward and reverse dynamics.

The probability to be in state $m$ at time $t$ under the reverse dynamics is written as $\rev{p}_m(t)$, or in vector notation $\rev{\mathbf{p}}_t$. The probability for a trajectory $\rev{\Pi}$ during the reverse dynamics and starting from $\rev{\mathbf{p}}_0$ is
\begin{multline}
\rev{\mathcal{P}}(\rev{\Pi}  \vert \rev{\mathbf{p}}_0)= \rev{p}_{m_N}(0)  e^{\int_{T-t_1}^{T} d \tau W_{m_{0},m_{0}}(\rev{\lambda}_{\tau})}\\
\times \left[ \prod_{j=1}^N e^{\int_{T-t_{j+1}}^{T-t_j} d \tau W_{m_{j},m_{j}}(\rev{\lambda}_{\tau})} W^{(\nu_j)}_{m_{j-1},m_j}(\rev{\lambda}_{T-t_j}) \right].
\end{multline}
The total trajectory EP is then defined as ($k_B=1$)
\begin{align}
\Delta s_{\mathrm{tot}} (\Pi\vert \mathbf{p}_0) &=  \ln  \frac{\mathcal{P}(\Pi  \vert \mathbf{p}_0)}{\rev{\mathcal{P}}(\rev{\Pi}\vert \mathbf{p}_T)} \label{eq::defstot} \\
&= \ln \frac{p_{m_0}(0)}{p_{m_N}(T)} + \sum_{j=1}^N \ln \frac{W^{(\nu_j)}_{m_j,m_{j-1}}(\lambda_{t_j})}{W^{(\nu_j)}_{m_{j-1},m_j}(\lambda_{t_j})} \label{eq::defstot2} \\
&= \Delta s_{\mathrm{sys}}(\Pi \vert \mathbf{p}_0) + \Delta s_{\mathrm{r}}(\Pi),
\end{align}
where the first term is the change in system entropy and the second term is the change in reservoir entropy. It is important to note that the end probability of the forward dynamics $\mathbf{p}_T=\mathbf{W}_{\mathrm{F}}\mathbf{p}_0$ is taken as the start probability of the time-reversed dynamics, i.e., $\rev{\mathbf{p}}_0= \mathbf{p}_T$.

As outlined in \cite{PRLesposito2010,PREesposito2010,PREvandenbroeck2010}, the total EP can be separated into adiabatic and non-adiabatic components. The adiabatic trajectory EP is defined as
\begin{align}
\Delta s_{\mathrm{a}} (\Pi) &=\sum^N_{j=1} \ln \frac{W^{(\nu_j)}_{m_j,m_{j-1}}(\lambda_{t_j})p^{\mathrm{s}}_{m_{j-1}}(\lambda_{t_j})}{W^{(\nu_j)}_{m_{j-1},m_{j}}(\lambda_{t_j})p^{\mathrm{s}}_{m_{j}}(\lambda_{t_j})} \label{eq::adiab1} \\
&=\sum^N_{j=1} \ln \frac{p^{\mathrm{eq},(\nu_j)}_{m_j}(\lambda_{t_j})p^{\mathrm{s}}_{m_{j-1}}(\lambda_{t_j})}{p^{\mathrm{eq},(\nu_j)}_{m_{j-1}}(\lambda_{t_j})p^{\mathrm{s}}_{m_{j}}(\lambda_{t_j})}\label{eq::adiab2},
\end{align}
where we used detailed balance:
\begin{equation}
W^{(\nu)}_{m_i,m_j}(\lambda) p^{\mathrm{eq},(\nu)}_{m_j}(\lambda)=W^{(\nu)}_{m_j,m_i}(\lambda) p^{\mathrm{eq},(\nu)}_{m_i}(\lambda).
\end{equation}
The non-adiabatic trajectory EP reads
\begin{align}
\Delta s_{\mathrm{na}} (\Pi\vert \mathbf{p}_0) &= \ln \frac{p_{m_0}(0)}{p_{m_N}(T)} +\sum^N_{j=1} \ln \frac{p^{s}_{m_{j}}(\lambda_{t_j})}{p^{s}_{m_{j-1}}(\lambda_{t_j})} \label{eq::nonad}.
\end{align}
The definitions of $\Delta s_{\mathrm{a}}$ and $\Delta s_{\mathrm{na}}$ are equivalent to \cite{PRLesposito2010}
\begin{align}
&\Delta s_{\mathrm{a}}(\Pi)=\ln \frac{\mathcal{P}(\Pi\vert \mathbf{p}_0)}{\mathcal{P}^+(\Pi\vert \mathbf{p}_0)},\\
&\Delta s_{\mathrm{na}}(\Pi\vert \mathbf{p}_0)=\ln \frac{\mathcal{P}(\Pi\vert \mathbf{p}_0)}{\rev{\mathcal{P}}^+(\rev{\Pi}\vert \mathbf{p}_T)},
\end{align}
where $+$ denotes that the system undergoes the adjoint dynamics with rates
\begin{equation}\label{eq::defadjoint}
W_{m,m'}^{(\nu)+}(\lambda_t) = \frac{W^{(\nu)}_{m',m}(\lambda_t) p^{s}_{m}(\lambda_t)}{p^{s}_{m'}(\lambda_t)}.
\end{equation}
The adjoint dynamics $\mathbf{W}^+(\lambda_t)$ has the same NESS $\mathbf{p}^{s}(\lambda_t)$ as the original dynamics $\mathbf{W}(\lambda_t)$. The dynamics $\mathcal{P}^+$ starts with probability distribution $\mathbf{p}_0$, while the dynamics $\rev{\mathcal{P}}^+$ starts with $\mathbf{p}_T$.\newline
The adiabatic EP is a measure of the difference between the instantaneous steady state $\mathbf{p}^{\mathrm{s}}(\lambda_t)$ and the equilibrium distributions $\mathbf{p}^{\mathrm{eq},(\nu)}(\lambda_t)$ of the reservoirs that cause the transitions. It is zero if the system is in contact with a single reservoir. Physically, it can be understood as the part of the total EP associated with nonequilibrium boundary conditions, i.e., coupling with different equilibrium reservoirs.
The non-adiabatic EP is zero if the system undergoes no external driving $\lambda_t = \lambda$, and if it starts in the steady state $\mathbf{p}_0 = \mathbf{p}^{\mathrm{s}}(\lambda)$. It is therefore seen as the part of the total EP associated with time-dependent driving and relaxation to the steady state.
\section{Three detailed fluctuation theorems}
Having defined the different EPs on the trajectory level, we now move on to the detailed fluctuation theorems which deal with probability distributions of the entropy production. We start by calculating the probability to observe a total EP $\Delta s_{\mathrm{tot}}$ in the forward dynamics, given the initial distribution $\mathbf{p}_0$. Using the definition Eq.~\eqref{eq::defstot} one finds:
\begin{align}
&P(\Delta s_{\mathrm{tot}}\vert \mathbf{p}_0) = \sum_{\Pi} \mathcal{P}(\Pi\vert \mathbf{p}_0) \delta \left( \Delta s_{\mathrm{tot}} - \ln  \frac{\mathcal{P}(\Pi\vert \mathbf{p}_0)}{\rev{\mathcal{P}}(\rev{\Pi}\vert \mathbf{p}_T)} \right) \label{epforw}\\
&\;\;= e^{\Delta s_{\mathrm{tot}}} \sum_{\Pi} \rev{\mathcal{P}}(\rev{\Pi}\vert \mathbf{p}_T) \delta \left( \Delta s_{\mathrm{tot}} - \ln  \frac{\mathcal{P}(\Pi\vert \mathbf{p}_0)}{\rev{\mathcal{P}}(\rev{\Pi}\vert \mathbf{p}_T)} \right) \\
&\;\;= e^{\Delta s_{\mathrm{tot}}} \sum_{\Pi} \rev{\mathcal{P}}(\Pi\vert \mathbf{p}_T) \delta \left( -\Delta s_{\mathrm{tot}} - \ln  \frac{\rev{\mathcal{P}}(\Pi\vert \mathbf{p}_T)}{\mathcal{P}(\rev{\Pi}\vert \mathbf{p}_0)} \right), \label{confuse}
\end{align}
with $\delta(\cdot)$ the Dirac $\delta$ function. The sum appearing on the right-hand side of Eq.~\eqref{confuse} is a normalized function with respect to the variable $\Delta s_{\mathrm{tot}}$, so it is tempting to regard it as the probability to observe a total EP $- \Delta s_{\mathrm{tot}}$ during the reverse dynamics. This is, however, not correct; the distribution for the total EP during the reverse dynamics, in analogy with Eq.~\eqref{epforw}, is
\begin{equation}
\rev{P}(\Delta s_{\mathrm{tot}} \vert \rev{\mathbf{p}}_0) = \sum_{\Pi}\rev{\mathcal{P}}(\Pi\vert \rev{\mathbf{p}}_0) \delta \left( \Delta s_{\mathrm{tot}} - \ln  \frac{\rev{\mathcal{P}}(\Pi\vert \rev{\mathbf{p}}_0)}{\mathcal{P}(\rev{\Pi}\vert \rev{\mathbf{p}}_T)} \right),
\end{equation}
where we have used that $\rev{\rev{\mathcal{P}}}(\rev{\Pi}\vert \rev{\mathbf{p}}_T) = \mathcal{P}(\rev{\Pi}\vert \rev{\mathbf{p}}_T)$, since $\rev{\rev{\lambda}}_t = \lambda_t$ by definition.
The initial distributions appearing inside the $\delta$ function are related via $\rev{\mathbf{p}}_T=\mathbf{W}_{\mathrm{R}}\rev{\mathbf{p}}_0$. It is clear that such a relation is not satisfied in general for Eq.~\eqref{confuse}, but only when $\mathbf{p}_0=\mathbf{W}_{\mathrm{R}}\mathbf{p}_T$. Hence only for initial conditions $\mathbf{p}_0$ satisfying the condition
\begin{equation}
\mathbf{p}_0 = \mathbf{W}_{\mathrm{R}}\mathbf{W}_{\mathrm{F}} \mathbf{p}_0. \label{echo}
\end{equation}
The requirement is that the end probability of the time-reversed process is equal to the start probability of the forward process. Initial conditions satisfying this requirement are called echo states, and are denoted by $\mathbf{p}^{\mathrm{echo}}_0$.
We can then write the detailed fluctuation theorem (DFT) for the total EP as follows:
\begin{equation}\label{eq::FT}
\frac{P(\Delta s_{\mathrm{tot}}\vert \mathbf{p}^{\mathrm{\tiny{echo}}}_0)}{\rev{P}(-\Delta s_{\mathrm{tot}}\vert \mathbf{W}_{\mathrm{F}}\mathbf{p}^{\mathrm{\tiny{echo}}}_0)}= e^{\Delta s_{\mathrm{tot}}}.
\end{equation}
If one starts from an echo state the entropy production is odd under time reversal [cf.~Eq.~\eqref{eq::defstot}]:
\begin{multline}\label{odd}
\Delta \rev{s}_{\mathrm{tot}} (\rev{\Pi}\vert \mathbf{W}_{\mathrm{F}}\mathbf{p}^{\mathrm{echo}}_0 )= \ln  \frac{\rev{\mathcal{P}}(\rev{\Pi}\vert \mathbf{W}_{\mathrm{F}}\mathbf{p}^{\mathrm{echo}}_0)}{\rev{\rev{\mathcal{P}}} ( \rev{\rev{\Pi}} \vert \mathbf{W}_{\mathrm{R}}\mathbf{W}_{\mathrm{F}}\mathbf{p}^{\mathrm{echo}}_0)} \\
=\ln  \frac{\rev{\mathcal{P}}(\rev{\Pi}\vert \mathbf{W}_{\mathrm{F}}\mathbf{p}^{\mathrm{echo}}_0)}{\mathcal{P}(\Pi  \vert \mathbf{p}^{\mathrm{echo}}_0)}= -\Delta s_{\mathrm{tot}} (\Pi\vert \mathbf{p}^{\mathrm{echo}}_0).
\end{multline}
The requirement Eq.~\eqref{echo} is equivalent to requiring that the EP is odd under time reversal for all paths $\Pi$. Indeed, from Eq.~\eqref{eq::defstot2} it is clear that the reservoir EP is always odd under time reversal: $\Delta s_{\mathrm{r}}(\Pi) = - \Delta \rev{s}_{\mathrm{r}}(\rev{\Pi})$. The system EP is odd under time reversal only if $\rev{p}_m(T) = p_m(0)$ for all $m$, i.e., if the initial condition is an echo state.

Echo states can be found by obtaining the eigenvector of $\mathbf{W}_{\mathrm{R}}  \mathbf{W}_{\mathrm{F}}$ with eigenvalue 1. Since $\mathbf{W}_{\mathrm{R}}  \mathbf{W}_{\mathrm{F}}$ is the product of two stochastic matrices, it is itself again a stochastic matrix. Hence there is at least one such eigenvector $\mathbf{p}^{\mathrm{echo}}_0$. If the matrix is furthermore irreducible and aperiodic, which we consider to be the typical case, the Perron-Frobenius theorem dictates that there is exactly one eigenvector with eigenvalue $1$.

We next turn to the DFTs for the adiabatic and non-adiabatic EP \cite{PRLesposito2010}. For the adiabatic EP we have
\begin{align}
&P(\Delta s_{\mathrm{a}}\vert \mathbf{p}_0) = \sum_{\Pi} \mathcal{P}(\Pi\vert \mathbf{p}_0) \delta \left( \Delta s_{\mathrm{a}} - \ln \frac{\mathcal{P}(\Pi\vert \mathbf{p}_0)}{\mathcal{P}^+(\Pi\vert \mathbf{p}_0)} \right)\nonumber\\
&=e^{\Delta s_{\mathrm{a}}}\sum_{\Pi} \mathcal{P}^+(\Pi\vert \mathbf{p}_0) \delta \left(-\Delta s_{\mathrm{a}} - \ln \frac{\mathcal{P}^+(\Pi\vert \mathbf{p}_0)}{\mathcal{P}(\Pi\vert \mathbf{p}_0)} \right)\nonumber \\
&=e^{\Delta s_{\mathrm{a}}}P^{+}(-\Delta s_{\mathrm{a}}\vert \mathbf{p}_0),
\end{align}
where we have used that $\mathcal{P}^{++}(\Pi\vert \mathbf{p}_0) = \mathcal{P}(\Pi\vert \mathbf{p}_0)$. Hence we can write
\begin{equation}
\frac{P(\Delta s_{\mathrm{a}}\vert \mathbf{p}_0)}{P^+(-\Delta s_{\mathrm{a}}\vert \mathbf{p}_0)} = e^{\Delta s_{\mathrm{a}}}, \label{eq::FTad}
\end{equation}
for \emph{any} initial distribution $\mathbf{p}_0$.

For the non-adiabatic EP we find
\begin{align}
&P(\Delta s_{\mathrm{na}}\vert \mathbf{p}_0) = \sum_{\Pi} \mathcal{P}(\Pi\vert \mathbf{p}_0) \delta \left( \Delta s_{\mathrm{na}} - \ln \frac{\mathcal{P}(\Pi\vert \mathbf{p}_0)}{\rev{\mathcal{P}}^+(\rev{\Pi}\vert \mathbf{p}_T)} \right)\nonumber\\
&=e^{\Delta s_{\mathrm{na}}}\sum_{\rev{\Pi}} \rev{\mathcal{P}}^+(\rev{\Pi}\vert \mathbf{p}_T) \delta \left(-\Delta s_{\mathrm{na}} - \ln \frac{\rev{\mathcal{P}}^+(\rev{\Pi}\vert \mathbf{p}_T)}{\mathcal{P}(\Pi\vert \mathbf{p}_0)} \right),
\end{align}
and for the time-reversed adjoint process:
\begin{multline}
\rev{P}^{+}(\Delta s_{\mathrm{na}}\vert \rev{\mathbf{p}}_0) =\\ \sum_{\rev{\Pi}} \rev{\mathcal{P}}^{+}(\rev{\Pi}\vert \rev{\mathbf{p}}_0) \delta \left( \Delta s_{\mathrm{na}} - \ln \frac{\rev{\mathcal{P}}^+(\rev{\Pi}\vert \rev{\mathbf{p}}_0)}{\mathcal{P}(\Pi\vert \mathbf{W}_{\mathrm{R},+}\rev{\mathbf{p}}_0)} \right),
\end{multline}
where we use that $\rev{\rev{\mathcal{P}}}^{+ +}(\Pi\vert \mathbf{W}_{\mathrm{R},+}\rev{\mathbf{p}}_0) = \mathcal{P}(\Pi\vert \mathbf{W}_{\mathrm{R},+}\rev{\mathbf{p}}_0)$, and where we have defined
\begin{equation}
\mathbf{W}_{\mathrm{R},+} = \left[ \overrightarrow{\exp} \int_0^T \mathbf{W}^+(\rev{\lambda}_t) dt \right].
\end{equation}
Initial conditions satisfying
\begin{equation}\label{eq::invcondnad}
\mathbf{p}_0 = \mathbf{W}_{\mathrm{R},+}  \mathbf{W}_{\mathrm{F}} \mathbf{p}_0
\end{equation}
are again called echo states, and are denoted as $\mathbf{p}^{\mathrm{echo}+}_0$.
The DFT for the non-adiabatic EP can be written as
\begin{equation}\label{eq::FTnad}
\frac{P(\Delta s_{\mathrm{na}}\vert \mathbf{p}^{\mathrm{echo+}}_0)}{\rev{P}^+(-\Delta s_{\mathrm{na}}\vert \mathbf{W}_{\mathrm{F}}\mathbf{p}^{\mathrm{echo}+}_0)} = e^{\Delta s_{\mathrm{na}}}.
\end{equation}
The derivation of this condition is completely analogous to the one for the total EP Eq.~\eqref{echo}.
For the echo states $\mathbf{p}^{\mathrm{echo}+}_0$, the non-adiabatic EP is odd under the adjoint time-reversed dynamics:
\begin{align}
\Delta \rev{s}^+_{\mathrm{na}} (\rev{\Pi}\vert \mathbf{W}_{\mathrm{F}}\mathbf{p}^{\mathrm{echo}+}_0) &= - \Delta s_{\mathrm{na}}(\Pi\vert \mathbf{p}^{\mathrm{echo+}}_0). \label{eq::conditionSna}
\end{align}
Starting from the echo state $\mathbf{p}^{\mathrm{echo}+}_0$ ensures that the system entropy is odd under the adjoint time-reversed dynamics, while the other term of the non-adiabatic EP is always odd under the adjoint time-reversed dynamics; see Eq.~\eqref{eq::nonad}. Equation \eqref{eq::invcondnad} is therefore equivalent to requiring that Eq.~\eqref{eq::conditionSna} holds for all paths $\Pi$. 

\section{Processes starting from the echo state}\label{sec::examples}

Consider a process $\lambda_t$ between $t=0$ and $t=T$. The echo state for the total EP can be calculated from Eq.~\eqref{echo}, with $\mathbf{W}_{\mathrm{F}}$ and $\mathbf{W}_{\mathrm{R}}$ given by Eq.~\eqref{eq::WFWR}. As such, $\mathbf{p}^{\mathrm{echo}}_0$ depends in an intricate manner on the dynamics of both the forward and the reverse processes. It is therefore difficult to make general comments on its properties. For some relevant special cases, the echo state can however be determined by a simple calculation.

An important class of processes which always start from the echo state for the total EP is nonequilibrium steady states \cite{PRLevans1993,PRLGallavotti1995,JPAkurchan1998,JCPgaspard2004,PRLSeifertNESS,PRLWang2002,PREWang2005,PREcleuren2006,EPLcleuren2007,PREcleuren2008,PRXkung,NPkoski}.
Since $\lambda_t = \lambda$ one has that $\rev{\lambda} = \lambda$ and $\mathbf{W}_{\mathrm{R}} = \mathbf{W}_{\mathrm{F}}$. Eq.~\eqref{echo} then reduces to $\mathbf{p}_0 = \exp \left[ \mathbf{W}(\lambda) 2 T \right] \mathbf{p}_0$, whose solution is $\mathbf{p}^{\mathrm{echo}}_0 = \mathbf{p}^{\mathrm{s}}(\lambda)$. In this case the non-adiabatic EP is zero, so the adiabatic and total EPs are equal. 

The echo state for the total EP can also be easily calculated for processes with periodic time-dependent rates that are symmetric under time reversal \cite{PRECrooks2008,PRLcleuren2006,CRPcleuren}. 
Consider a system subject to a periodic driving $\lambda_{t+\tau} = \lambda_t$, with $\tau$ the period. In the long-time limit the system is in a time-dependent periodic steady state $\mathbf{p}^{\mathrm{ps}}(\lambda_t)$ \cite{JSPshargel}:
\begin{equation}\label{eq::expss}
\mathbf{p}^{\mathrm{ps}}(\lambda_{t+\tau}) = \mathbf{p}^{\mathrm{ps}}(\lambda_t).
\end{equation}
Consider the situation where $T = \tau$, with a driving symmetric under time reversal: $\rev{\lambda}_t = \lambda_{\tau-t} = \lambda_t$.
In this case $\mathbf{W}_{\mathrm{R}} = \mathbf{W}_{\mathrm{F}}$ and $\mathbf{p}^{\mathrm{ps}}(\lambda_0)$ is the echo state:
\begin{align}
\mathbf{W}_{\mathrm{F}} \mathbf{W}_{\mathrm{F}} \mathbf{p}^{\mathrm{ps}}(\lambda_0) =  \mathbf{p}^{\mathrm{ps}}(\lambda_{2\tau}) = \mathbf{p}^{\mathrm{ps}}(\lambda_0).
\end{align}
The DFT for the total EP has been verified experimentally for this situation \cite{PRLSeifert,PRLBlickle2006,EPLjoubaud2008}. We stress that $\mathbf{p}^{\mathrm{ps}}(\lambda_0)$ is, in general, not the echo state for the non-adiabatic EP. 

In the limit $t \uparrow \infty$ the contribution of the initial condition to the EPs can become negligible. More precisely, the conditions Eqs.~\eqref{odd} and \eqref{eq::conditionSna} are violated because of the contribution of the system entropy in respectively Eqs.~\eqref{eq::defstot2} and \eqref{eq::nonad}. The reservoir EP typically grows with time. Hence, when the system EP is bounded, its contribution becomes negligible in the limit $t \uparrow \infty$. In this limit one recovers the asymptotic fluctuation for the total EP \cite{PRLGallavotti1995,JSPLebowitz1999,JSPmaes}. If the state space is infinite, the asymptotic fluctuation theorem can be invalid \cite{JSMharris}.

Finally, we mention a particular scenario for the total EP described in \cite{PREcuetara} which reproduces the echo state. A system in contact with several reservoirs is prepared so that its initial state is the equilibrium distribution of one particular reservoir at $\lambda_0$. When the forward process is finished, the system is allowed to relax to the same equilibrium distribution but now at $\lambda_T$. This distribution is the start for the reverse process, after which the system relaxes again to the initial equilibrium distribution at $\lambda_0$.

\section{Echo states for a modulated quantum dot}
\begin{figure}[t]
\centering
\includegraphics[width=0.6\columnwidth]{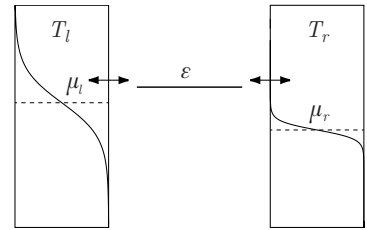}
\caption{Sketch of the model: a single energy level with energy $\epsilon$, connected to two electron reservoirs at different chemical potentials and temperatures.}
\label{fig::qd}
\end{figure}

\begin{figure}
\centering
\includegraphics[width=1.0\columnwidth]{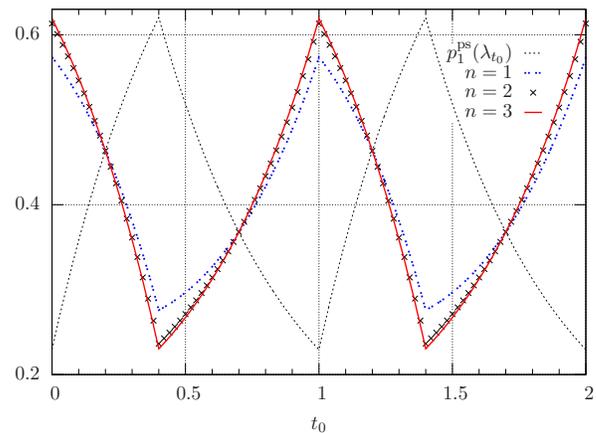}
\caption{(color online). $p_1^{\mathrm{echo}}(t_0,n)$ for $n=1$ (blue double dotted line), $n=2$ (black crosses), $n=3$ (red line), and $p_1^{\mathrm{ps}}(\lambda_{t_0})$ (black dotted line).
}
\label{fig::pEvVSpLim}
\end{figure}

\begin{figure}
\centering
\includegraphics[width=1.0\columnwidth]{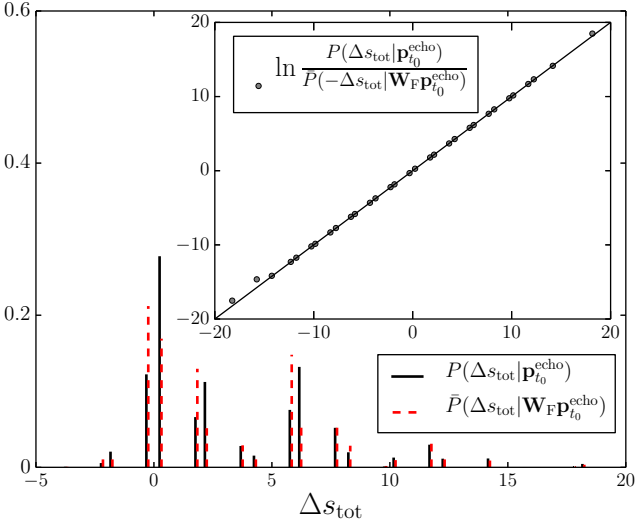}
\caption{(color online). Probability distributions for the total EP for the forward and time-reversed dynamics, starting from, respectively, $\mathbf{p}^{\mathrm{echo}}_{t_0}$ and $\mathbf{W}_{\mathrm{F}} \mathbf{p}^{\mathrm{echo}}_{t_0}$. The verification of the DFT for the total EP is shown in the inset.}
\label{fig::check_FT}
\end{figure}

\begin{figure}
\centering
\includegraphics[width=1.0\columnwidth]{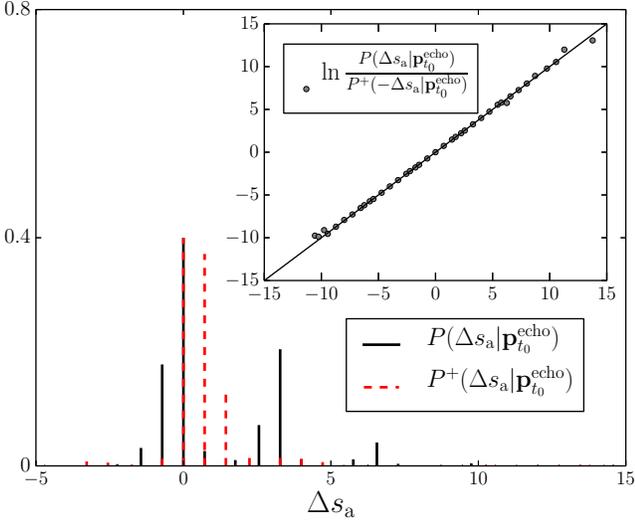}
\caption{(color online). Probability distributions for the adiabatic EP for the forward and adjoint dynamics, both starting from $\mathbf{p}^{\mathrm{echo}}_{t_0}$. The verification of the DFT for the adiabatic EP is shown in the inset.}
\label{fig::check_FT_ad}
\end{figure}

\begin{figure}
\centering
\includegraphics[width=1.0\columnwidth]{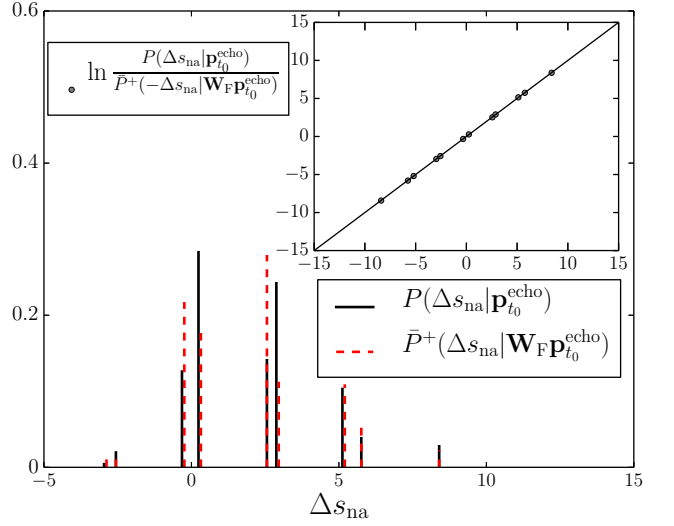}
\caption{(color online). Probability distributions for the non-adiabatic EP for the forward and reverse adjoint dynamics, starting from, respectively, $\mathbf{p}^{\mathrm{echo}}_{t_0}$ and $\mathbf{W}_{\mathrm{F}} \mathbf{p}^{\mathrm{echo}}_{t_0}$. The verification of the DFT for the non-adiabatic EP is shown in the inset.}
\label{fig::check_FT_nad}
\end{figure}

\begin{figure}
\centering
\includegraphics[width=0.8\columnwidth]{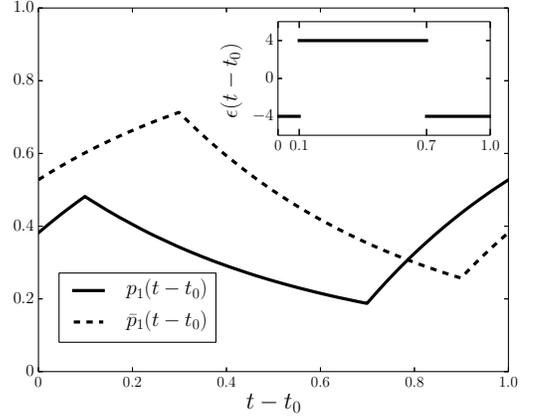}
\caption{Probabilities to be occupied $p_1(t-t_0)$ and $\rev{p}_1(t-t_0)$, starting from, respectively, $\mathbf{p}^{\mathrm{echo}}_{t_0}$ and $\mathbf{W}_{\mathrm{F}} \mathbf{p}^{\mathrm{echo}}_{t_0}$, with $t_0 = 0.3$. The time dependence of the energy level is shown in the inset.}
\label{fig::pt_involution}
\end{figure}

We illustrate our results on a modulated quantum dot. The stochastic thermodynamics of this model has already been discussed for various modes of operation, cf.~\cite{PRB2006,EPL2009,PRE2010,EPJBWillaert}. The model consists of a quantum dot with a single energy level exchanging electrons with two reservoirs; see Figure \ref{fig::qd}. The energy level is either empty (0) or occupied by a single electron (1). The transition rates are
\begin{align}
W^{(\nu)}_{10}(\lambda_t) = a_{\nu} f( x^{\nu}_t ), \quad W^{(\nu)}_{01}(\lambda_t) = a_{\nu} \left[ 1 - f(x^{\nu}_t)  \right],
\end{align}
where $\nu$ denotes the left $(l)$ or right $(r)$ reservoir, $f(x) = \left[ \exp(x) + 1 \right]^{-1}$ is the Fermi distribution, and $a_{\nu}$ is the system-reservoir coupling. The chemical potential and temperature of the reservoirs are denoted by, respectively, $\mu_{\nu}$ and $T_{\nu}$, and the control variable is the value of the energy level $\lambda_t = \epsilon_t$. The variable in the Fermi distribution is $x^{\nu}_t = \left( \lambda_t - \mu_{\nu} \right) / T_{\nu}$.

We consider a piecewise constant periodic driving of the form:
\begin{equation}
\lambda_t =
\begin{cases}
\epsilon_I, & 0 < t \bmod{\tau} \leq \alpha \tau \\
\epsilon_{II}, & \alpha \tau  < t \bmod{\tau} \leq \tau,
\end{cases}
\end{equation}
where $\epsilon_I$ and $\epsilon_{II}$ are constants, $0 \leq \alpha \leq 1$, and $\tau$ is the period. Since for this model $\mathbf{W}(\lambda_t) = \mathbf{W}^+(\lambda_t)$, both echo states $\mathbf{p}^{\mathrm{echo}}$ and $\mathbf{p}^{\mathrm{echo}+}$ are identical. The forward dynamics is run over $n$ periods, with $n$ an integer. The start time is denoted by $t_0 = \alpha_0 \tau$, with $0 \leq \alpha_0 \leq 1$. The echo state is written as follows: 
\begin{equation}
\mathbf{p}^{\mathrm{echo}} = \left\{p^{\mathrm{echo}}_0 , p^{\mathrm{echo}}_1  \right\} = \left\{\frac{1}{1+y} , \frac{y}{1+y}  \right\}.
\end{equation}
By identification with the eigenvector with eigenvalue $1$ in Eq.~\eqref{echo} or \eqref{eq::invcondnad}, one finds the explicit expression:
\begin{align}\label{eq::invcondexactNT}
y= \begin{cases}
   z(\alpha_0,\alpha,w_I,w_{II},n),& \alpha_0 \leq \alpha, \\
   z(\alpha_0-\alpha,1-\alpha,w_{II},w_I,n),& \alpha_0 > \alpha,
  \end{cases}
\end{align}
with $a = a_l + a_r$, $w_I = W_{10}(\epsilon_I)$, $w_{II} = W_{10}(\epsilon_{II})$, and
\begin{align}
&z(\alpha_0,\alpha,w_I,w_{II},n) \notag \\ 
&= \frac{w_I \sinh(n a \tau)+(w_I-w_{II}) g(\alpha_0,\alpha,n)}{(a-w_I)\sinh(n a \tau)-(w_I-w_{II}) g(\alpha_0,\alpha,n)}, \\
&g(\alpha_0,\alpha,n) \notag \\
&= \sum^{n-1}_{k=0}\left[\sinh(a \tau (k+\alpha_0))-\sinh(a \tau (k+\alpha_0+1-\alpha) )\right].\label{eq::invcondfN} 
\end{align}
Eq.~\eqref{eq::invcondexactNT} was checked analytically for $n=1,2,3$ for all parameter values, and numerically for $n > 3$ for different sets of parameter values. 

From here on we consider the following choice of  parameters: $\mu_l = 2, \mu_r = -2, a_l = a_r = 1, T_l = T_r = 1, \epsilon_I = -4, \epsilon_{II} = 4, \alpha = 0.4,$ and $\tau = 1$.
We plot in Figure \ref{fig::pEvVSpLim} $p_1^{\mathrm{echo}}(t_0,n)$ as a function of $t_0$ for $n= 1, 2,$ and $3$. It is clearly very different from the periodic steady state $p_1^{\mathrm{ps}}(\lambda_{t_0})$.
In the limit of large $n$, one finds from Eq.~\eqref{eq::invcondexactNT} that: 
\begin{equation}
\lim_{n \rightarrow \infty} \mathbf{p}^{\mathrm{echo}}(t_0,n) = \rev{\mathbf{p}}^{\mathrm{ps}}(\rev{\lambda}_{\tau-t_0}).
\end{equation}
For large $n$ the time-reversed process is in its periodic steady state $\rev{\mathbf{p}}^{\mathrm{ps}}(\rev{\lambda}_t)$, where $\rev{\lambda}_{t} = \lambda_{\tau - t}$. If one starts from $t_0$ the final distribution of the time-reversed process is, in this limit, equal to $\rev{\mathbf{p}}^{\mathrm{ps}}(\rev{\lambda}_{\tau-t_0})$. The echo state is therefore equal to this distribution.
Note that $\mathbf{p}^{\mathrm{ps}}(\lambda_{t_0})$ coincides with $\mathbf{p}^{\mathrm{echo}}(t_0, n)$ for $t_0 = 0.2$ and $t_0 = 0.7$. For these start times the driving is symmetric under time reversal. In this case the model falls under the ``trivial'' category of time-symmetric drivings discussed in Section \ref{sec::examples}.
 
Having identified the echo state, we determined the various entropy productions via numerical simulations using the algorithm from \cite{EPLholubec2011}, for the specific choice $t_0 = \alpha_0 \tau = 0.3$ and $n = 1$. The thus obtained distributions $P(\Delta s_{\mathrm{tot}}\vert \mathbf{p}^{\mathrm{echo}}_{t_0})$ and $\rev{P}(\Delta s_{\mathrm{tot}}\vert \mathbf{W}_{\mathrm{F}} \mathbf{p}^{\mathrm{echo}}_{t_0})$ are shown in Figure \ref{fig::check_FT}. The two $\delta$ peaks of both distributions around $\Delta s_{\mathrm{tot}} = 0$ are for the trajectories that have no transition. The four other large $\delta$ peaks are for trajectories with one transition. The DFT is satisfied; cf.~the inset of Figure \ref{fig::check_FT}.
$P(\Delta s_{\mathrm{a}}\vert \mathbf{p}^{\mathrm{echo}}_{t_0})$ and $P^+(\Delta s_{\mathrm{a}}\vert \mathbf{p}^{\mathrm{echo}}_{t_0})$ are represented in Figure \ref{fig::check_FT_ad}, and $P(\Delta s_{\mathrm{na}}\vert \mathbf{p}^{\mathrm{echo}}_{t_0})$ and $\rev{P}^+(\Delta s_{\mathrm{na}}\vert \mathbf{W}_{\mathrm{F}} \mathbf{p}^{\mathrm{echo}}_{t_0})$ in Figure \ref{fig::check_FT_nad}. The DFTs Eqs.~\eqref{eq::FTad} and \eqref{eq::FTnad} are both satisfied, see the insets. 
The probabilities $p_1(t)$ and $\rev{p}_1(t)$ starting from, respectively, $\mathbf{p}^{\mathrm{echo}}_{t_0}$ and $\mathbf{W}_{\mathrm{F}} \mathbf{p}^{\mathrm{echo}}_{t_0}$ are shown in Figure \ref{fig::pt_involution}.

\section{Shadowing the Echo States}

\subsection{Total entropy production}

Suppose one wants to produce experimentally the echo state for the total EP, for the driving $\lambda_t$ between $t=0$ and $t=T$. This can be done by applying the following driving $\lambda'_t$ to the system:
\begin{equation}\label{eq::shadowecho}
\lambda'_t =
\begin{cases}
\lambda_t, & 0 < t \bmod{2T} \leq T, \\
\rev{\lambda}_t, & T < t \bmod{2T} \leq 2T.
\end{cases}
\end{equation}
As is specified by the modulo $2T$ prescription, this driving is periodic with period $\tau = 2T$. The echo state is the periodic steady state at $\lambda'_0$: $\mathbf{p}^{\mathrm{echo}}_0 = \mathbf{p}^{\mathrm{ps}}(\lambda'_0)$. 

It is, however, not necessary to prepare the system in the echo state. It is well known that one can reconstruct probability distributions from measurements under a different distribution, for example, via umbrella sampling \cite{frenkel2001}. We introduce here a procedure that, starting from any initial condition, reproduces the distribution of the total entropy production when starting from the echo state. This procedure could be applied to already existing experimental data.

Consider a collection of experimentally measured paths $\{ \Pi \}$ ($\{ \rev{\Pi} \}$) starting from some arbitrary initial distribution $\mathbf{p}_0$ ($\mathbf{\rev{p}}_0$), measured under the forward (reverse) dynamics.
One can find $\mathbf{W}_{\mathrm{F}}$ and $\mathbf{W}_{\mathrm{R}}$ from, respectively, $\{ \Pi \}$ and $\{ \rev{\Pi} \}$, at least if all transitions are possible ($p_m(0) \neq 0$ and $\rev{p}_m(0) \neq 0$ for all $m$). The transition matrix $\mathbf{W}_{\mathrm{R}} \mathbf{W}_{\mathrm{F}}$ can then be used to find the echo state $\mathbf{p}^{\mathrm{echo}}_0$. Suppose now one has measured the reservoir entropies $\Delta s_{\mathrm{r}} (\Pi)$, which are independent of the starting probability. 
The trajectory entropies starting from the echo state are found by:
\begin{align}
\Delta s_{\mathrm{tot}} (\Pi\vert \mathbf{p}^{\mathrm{echo}}_0) = \Delta s_{\mathrm{r}}(\Pi) + \ln \frac{p^{\mathrm{echo}}_{m_0}(0)}{p^{\mathrm{echo}}_{m_N}(T)}. \label{eq::corrinvstot}
\end{align}
If instead one has measured the total EP, the original system EP $\ln p_{m_0}(0) / p_{m_N}(T)$ must be subtracted in Eq.~\eqref{eq::corrinvstot}, where $\mathbf{p}_0$ can be found from the collection of paths $\{ \Pi \}$. The corrected EPs from Eq.~\eqref{eq::corrinvstot} can be used to create the probability distributions for the total EP when starting from the echo state as follows. 
Consider each collection of paths that start with the same state $m_0$ separately. For each such collection, calculate the probability distribution for the EPs of Eq.~\eqref{eq::corrinvstot}. These probability distributions are denoted by $P_{m_0}(\Delta s_{\mathrm{tot}} \vert \mathbf{p}^{\mathrm{echo}}_0)$. The probability distribution of the total EP when starting from the echo state is then found by:
\begin{equation}
P(\Delta s_{\mathrm{tot}} \vert \mathbf{p}^{\mathrm{echo}}_0) = \sum_{m_0} p^{\mathrm{echo}}_{m_0}(0) P_{m_0}(\Delta s_{\mathrm{tot}} \vert \mathbf{p}^{\mathrm{echo}}_0).
\end{equation}
A completely analogous procedure can be followed for the time-reversed process. Figure \ref{fig::check_FT} was reproduced with this procedure, for $p_1(0) = 0.5$ instead of $p^{\mathrm{echo}}_1(0) = 0.3808$.

\subsection{Non-adiabatic entropy production}

The echo state for the non-adiabatic EP can be found by producing the periodic steady state of the dynamics where the evolution over each period is described by $\mathbf{W}_{\mathrm{R},+} \mathbf{W}_{\mathrm{F}}$. The experimental realization of the $\mathbf{W}_{\mathrm{R},+}$ dynamics is in general not a trivial exercise, since all transition probabilities $W^{(\nu)}_{m,m'}(\lambda_t)$ have to be separately changed according to Eq.~\eqref{eq::defadjoint}. For the modulated quantum dot the adjoint dynamics is readily obtained, since one only has to change the chemical potentials of the reservoirs. We are not aware of a general scheme to produce the adjoint dynamics experimentally, given the original dynamics. Since the adjoint dynamics is needed for both the adiabatic and non-adiabatic DFT, this is an interesting question for further research.

\section{Conclusion}

As was pointed out by Seifert \cite{PRLSeifertNESS}, the inclusion of the so-called stochastic system entropy production allows one to derive integral fluctuation theorems valid for finite times. The situation is more delicate for the detailed fluctuation theorems. If one wants to interpret the quantities associated to the reverse (reverse adjoint) process as the total (non-adiabatic) entropy of that process, one needs to make a specific choice of the initial condition, which is typically unique. For these so-called echo states, the starting probability distribution of the original dynamics and final probability distribution of the transformed dynamics are equal. Starting from an echo state ensures that the system entropy is odd under the transformed dynamics. As a result, both the total and non-adiabatic entropy productions are odd under their respective transformed dynamics; cf.~Eqs.~\eqref{odd} and \eqref{eq::conditionSna}.
Stochastic quantities such as heat, work, and entropy production have by now been measured experimentally in a wide variety of systems. Our prescriptions should thus be easily verifiable, either by choosing the echo states as proper initial conditions in the experiments, or by applying our shadowing operation when starting from other initial states that are more easily implemented, such as a long-time periodic steady state.

\begin{acknowledgments}
Preliminary work was performed by Cedric Driesen. This work was supported by the Flemish Science Foundation (Fonds Wetenschappelijk Onderzoek). The computational resources and services used in this work were provided by the VSC (Flemish Supercomputer Center), funded by the Hercules Foundation and the Flemish Government, department EWI.
\end{acknowledgments}


\end{document}